\documentclass[aps,prr,reprint,twocolumn,superscriptaddress,floatfix,amsmath, amssymb, longbibliography]{revtex4-2}
\usepackage[utf8]{inputenc}
\usepackage{graphicx}
\usepackage{soul} 
\usepackage{dcolumn}
\usepackage{bm}
\usepackage{color}

\usepackage{graphicx}
\usepackage{dcolumn}
\usepackage{bm}
\usepackage{amssymb}
\usepackage{amsmath}
\usepackage{multirow}
\usepackage{color}
\usepackage{hyperref}
\usepackage{float}

\def\spose#1{\hbox to 0pt{#1\hss}}
\def\lesssim{\mathrel{\spose{\lower 3pt\hbox{$\mathchar"218$}}
  \raise 2.0pt\hbox{$\mathchar"13C$}}}
\def\gtrsim{\mathrel{\spose{\lower 3pt\hbox{$\mathchar"218$}}
  \raise 2.0pt\hbox{$\mathchar"13E$}}}

\begin{document}
\title{Nonequilibrium thermodynamics of DNA nanopore unzipping}
\date{\today}


\author{Antonio Suma}
\affiliation{Dipartimento di Fisica, Universit\'a degli Studi di Bari and INFN, Sezione di Bari, via Amendola 173, Bari, I-70126, Italy}
\affiliation{Institute for Computational Molecular Science, Temple University, Philadelphia, PA 19122, USA }
\author{Vincenzo Carnevale}
\affiliation{Institute for Computational Molecular Science, Temple University, Philadelphia, PA 19122, USA }
\author{Cristian Micheletti}
\affiliation{Scuola Internazionale Superiore di Studi Avanzati (SISSA), Via Bonomea 265, 34136 Trieste, Italy}

\begin{abstract}
Using theory and simulations, we carried out a first systematic characterization of DNA unzipping via nanopore translocation. Starting from partially unzipped states, we found three dynamical regimes depending on the applied force, $f$: (i) heterogeneous DNA retraction and rezipping  ($f < 17$pN), (ii) normal ($17$pN $< f < \textcolor{black}{60}$pN) and (iii) anomalous ($f > \textcolor{black}{60}$pN) drift-diffusive behavior. We show that the normal drift-diffusion regime can be effectively modelled as a one-dimensional stochastic process in a tilted periodic potential. \textcolor{black}{We use the theory of stochastic processes to recover the potential from nonequilibrium unzipping trajectories and show that it corresponds to the free-energy landscape for single base-pairs unzipping. Applying this general approach to other single-molecule systems with periodic potentials ought to yield detailed free-energy landscapes from out-of-equilibrium trajectories.}
\end{abstract}

\maketitle

Polymer translocation, the process of forcing polymers through narrow pores, has long been studied for its complex nonequilibrium properties\cite{chuang2001anomalous,kantor2004anomalous,palyulin2014polymer,keyser2021}, as well as for its practical relevance for molecular sensing\cite{liu2019detecting,chen2017direction,dorfman2014hydrodynamics}, label-free DNA sequencing~\cite{branton2008potential,tsutsui2010identifying,gundlach2012,choudhary2020high} and for \textit{in vivo} transactions of DNA and RNAs~\cite{bustamante2003ten,pyle2008,akiyama2016zika,suma2020directional}.
Various theoretical and experimental breakthroughs have illuminated crucial aspects of the process, from the propagation of mechanical tension along the polymer contour~\cite{kantor2004anomalous,muthukumar2007mechanism,sakaue2007nonequilibrium,ikonen2012unifying,rowghanian2012propagation,palyulin2014polymer,keyser2021} and the induced unraveling of the folds~\cite{grosberg2006long,keyser2021} to the hindrance or friction introduced by topological entanglements~\cite{suma2015pore,plesa2016direct,suma2017pore,sharma2019complex,suma2020directional}.
Scaling arguments~\cite{kantor2004anomalous,grosberg2006long,sakaue2007nonequilibrium} and stochastic modeling~\cite{muthukumar2007mechanism,ikonen2012unifying} have further helped rationalize the effects on translocation of intrinsic polymer properties such as thickness, contour length, and flexibility and external factors, e.g.~pore size and magnitude of the driving force.

The important case where the polymer structure is itself altered by translocation is, however, still underexplored. Ubiquitous examples are found in living cells, where double-stranded DNA and folded RNAs are translocated and unzipped by ATP-fueled enzymes~\cite{pyle2008}. Nanopore-based DNA unzipping is essential in genome sequencing setups~\cite{gundlach2012,Loose2018} and DNA capture and threading processes, too\cite{branton2003,meller2004,meller2008,comer2009microscopic}. \textcolor{black}{Notably, unzipping experiments combined with advanced theoretical methods allowed to characterize the abrupt unzipping transition thermodynamics of RNA motifs or DNA hairpins of tens of base pairs\cite{dudko2007,dudko2008theory,gupta2011experimental}.}

\textcolor{black}{Despite these efforts, we still lack a comprehensive and predictive theory for DNA nanopore unzipping dynamics and for using it to extract detailed thermodynamic information. Specifically, no single-base-pair free energy profiles have yet been derived from nonequilibrium unzipping trajectories. Thus, two fundamental questions are still unanswered:} how do the highly dissipative unzipping steps affect the translocation dynamics? Can we glean information about the thermodynamics of these microscopic events from the sole observation of nanopore translocation trajectories?

Here, using theory and computation, we study the pore unzipping process of DNAs of hundreds of base pairs and address the nonequilibrium thermodynamics in two stages. First, we characterize how the process depends on the driving force, and discuss the results with reference to known types of stochastic processes. Second, by harnessing the nonequilibrium translocation traces, we infer the potential of mean force, or thermodynamic landscape, of the microscopic DNA unzipping steps. This observation suggests a novel strategy to glean thermodynamic information from nonequilibrium trajectories and solve this inverse problem for a wide range of periodic systems with drift-diffusion, including complex biomolecular ones\cite{hayashi2015giant,kim2017giant}.

\begin{figure*}[t]
\centering 
\includegraphics[width=1.8\columnwidth]{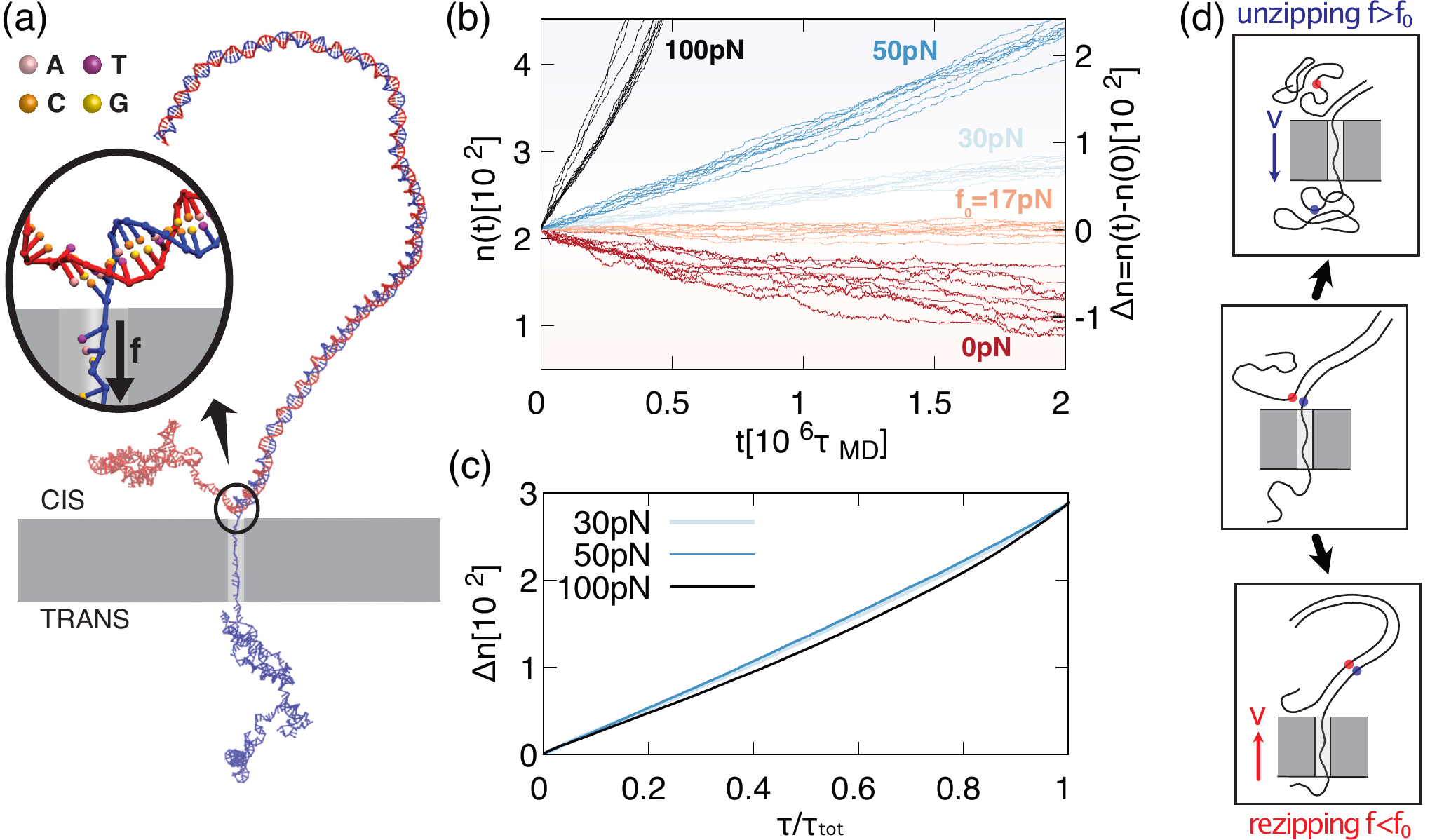}
\caption{(a) System setup. Typical configuration of a DNA chain of 500 base-pairs (oxDNA2 model) as it unzips during the driven pore translocation of one of its strands. Translocation is driven by the external force, $f$, which exclusively acts on the DNA tract in the pore. A magnified view of the pore entrance is shown in the inset, where the random DNA composition is color-coded as in the legend. (b) Translocation traces. The traces show the time evolution of the index of the nucleotide at the pore entrance, $n(t)$, for different driving forces, $f$. The traces of ten individual trajectories are shown for each force. Starting from a partially-unzipped state, $n(0)\sim 220$, translocation proceeds forward (unzipping) or backward (rezipping) depending on the driving force magnitude, see sketches in panel (d). The number of translocated nucleotides since the start of the simulation is thus $\Delta n(t) = n(t) -n(0)$. (c) The traces show the average translocation time $\tau$ ($x$-axis) of $\Delta n$ nucleotides ($y$-axis) at different forces, normalised to the average translocation duration, $\tau_{\rm tot}$.}
\label{fig1}
\end{figure*}

\textit{Model and system setup.} We considered DNA duplexes of 500 random base pairs (bp) driven through a narrow cylindrical pore running perpendicularly through a slab, see Fig.~\ref{fig1}. The pore width and slab thickness were set to $1.87$nm and $8.52$nm, respectively, as typical for nanopores~\cite{branton2003,meller2004,meller2008}. The pore lumen can thus accommodate a single-stranded DNA stretch of about 14 nucleotides at an average longitudinal spacing of 0.60nm.

To describe the duplex we used the oxDNA2 model\cite{oxdna1,oxdna2,oxdna3}, which accounts for sequence specificity and adopts three interaction centers per nucleotide with excluded volume, base-pairing, stacking, and screened electrostatic interactions, see Supplemental Material (SM) for further details. \footnote{The Supplemental Material is provided at [URL will be inserted by publisher] and includes additional Ref.~\cite{suma2019tacoxdna}}. The Debye screening length was set to 0.3nm, appropriate for 1M monovalent salt solutions. A truncated and shifted Lennard-Jones potential was used for the excluded-volume interactions of the DNA with the slab and pore walls, see SM. Translocation was driven by a longitudinal force, $f$, acting exclusively on the DNA tract inside the pore. \textcolor{black}{Because the length of the latter fluctuates in time, $f$ was equally subdivided among the nucleotides instantaneously inside the pore, a technical expedient to work at constant total driving force.}

The Langevin dynamics of the system at $T=300K$ was integrated with the LAMMPS simulation package~\cite{LAMMPS,henrich2018} with a timestep of $0.01\tau_{MD}$, $\tau_{MD}$ being the characteristic simulation time. The unitless parameters for mass, $m=3.15$, and friction, $\gamma=5$, were set to ensure that the inertial contribution to translocation is negligible at all considered forces, see SM. The partially unzipped initial configurations were obtained by translocating the first 200 nucleotides of equilibrated DNA duplexes, see Fig.~\ref{fig1}. From 20 to 80 independent translocation trajectories were carried out at different forces, randomly varying the DNA composition for each run.

\textit{Dynamical regimes.} The translocation response of the system is overviewed in Fig.~\ref{fig1}b where ten traces per force, corresponding to as many trajectories, illustrate the time-evolution of the number of translocated bases since the start of the simulations, $\Delta n(t)$. The dynamics of the entire unzipping process is recapitulated in Fig.~\ref{fig1}c, which shows the normalized average translocation time ($x$-axis) of increasing numbers of nucleotides ($y$-axis) at various driving forces. The translocation dynamics is further detailed in the SM.

Two key facts emerge from the inspection of the traces. First, the partially unzipped strand retracts from the pore at small applied forces. The retraction is caused by the rezipping of the unpaired bases at the pore entrance. The rezipping opposes the driven translocation with a force of $f_0\sim$17pN, which is the magnitude of the required force to stall the process. Taking into account the typical longitudinal base spacing of 0.6nm, one thus has that the unzipping free-energy cost per base pair is $10.2$pN\,nm on average, which compares well with the value of 11.15 pN nm from consensus thermodynamic parameters\cite{SantaLucia}. The observed stalling force is consistent with the one reported in DNA unzipping experiments with optical tweezers\cite{dieterich2016control}, which was approximately 17pN, too.

\begin{figure}[h]
\centering 
\includegraphics[width=0.98\columnwidth]{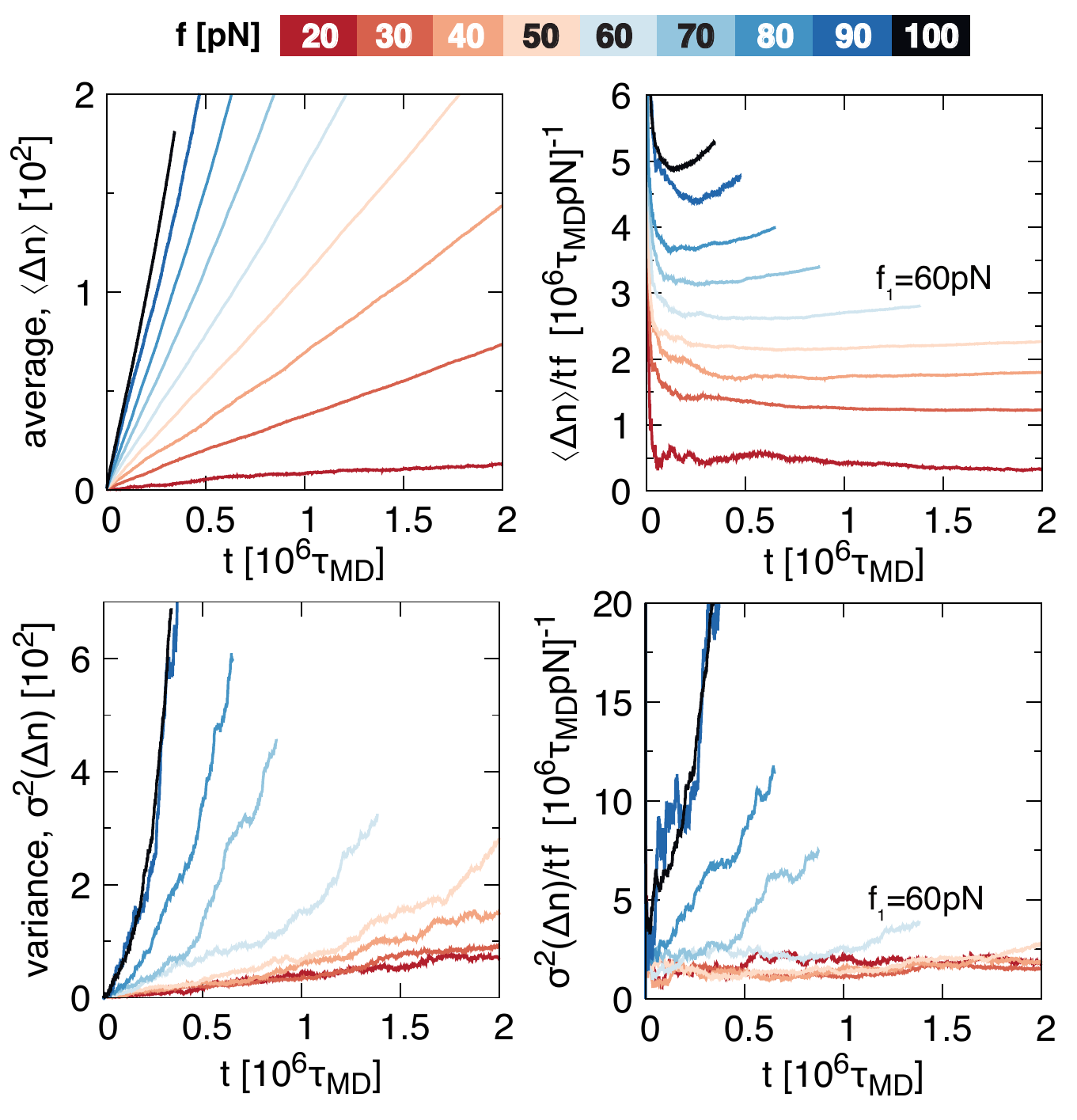}
\caption{Drift-diffusion regimes. For  $ f_0 <  f< f_1 \sim \textcolor{black}{60}$pN, the average and the variance of the number of translocated nucleotides, $\langle \Delta n \rangle$ and $\sigma^2({\Delta n})$, have an overall linear time dependence, as normal drift-diffusive processes. The crossover to the anomalous regime for $f>f_1$ is best appreciated in the right panels,  where $\langle \Delta n \rangle$ and $\sigma^2({\Delta n})$ are divided by $t$ and, additionally, by $f$, to facilitate comparison on a similar scale.}
\label{fig2}
\end{figure}

The rezipping process can be highly heterogeneous, as revealed by the spread of individual traces. In fact, each of the two unzipped strands is prone to the stochastic formation of non-native base pairings, and those formed in the {\em trans} side inevitably hinder the backward translocation, see SM. Second, in the forward-directed process ($f> f_0$), the $n(t)$ traces change from an overall linear to non-linear time-dependence at sufficiently large forces.
This crossover occurs at $f =f_1\sim \textcolor{black}{60}$pN, as detailed in Fig.~\ref{fig2}. The left panels in the figure present the time evolution of the mean and variance of  $\Delta n(t)$. The right panels show the same data divided by time and force, for an easier comparison at a similar scale. For $f < f_1$, the time-rescaled curves of the mean $\langle \Delta n(t)\rangle/(t \, f) $ and the variance $\sigma^2 (\Delta n(t)) / (t\, f)$ are both flat, as in normal drift-diffusive processes. At larger forces, instead, the non-trivial time dependence of the same two quantities indicates a crossover towards an anomalous drift-diffusion regime.

We thus conclude that DNA unzipping by translocation can be assimilated to a drift-diffusive process for $f_0$ $\lesssim f \lesssim f_1$. This result, with far-reaching implications as discussed below, has no analog in conventional DNA translocation (without the dissipative unzipping process) as it is not homogeneous in time because of the mechanical tension generated at the pore and propagating along the rest of the DNA contour.

\begin{figure}[tb]
\centering 
\includegraphics[width=0.4\textwidth]{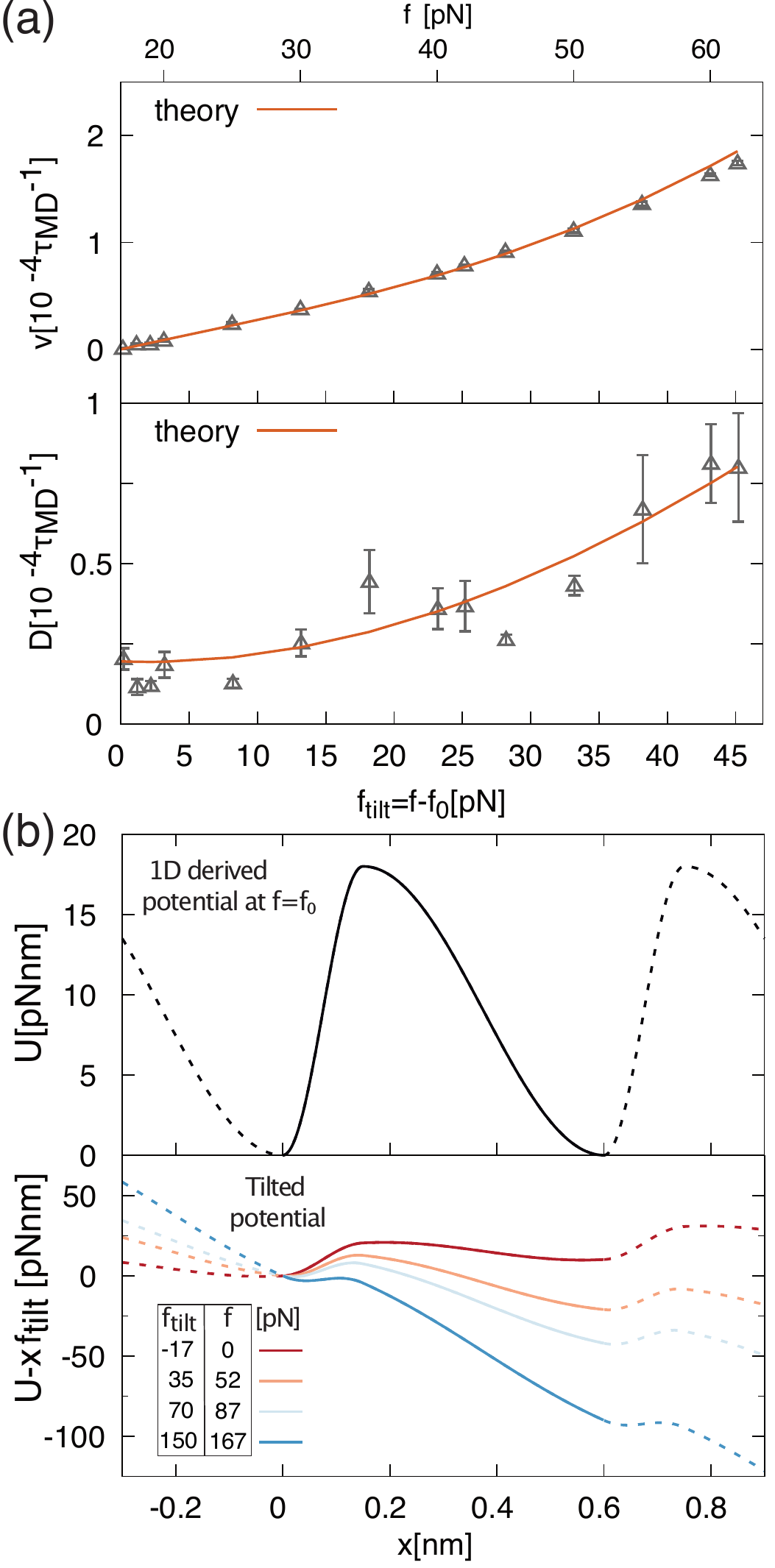}
\caption{{Derivation of the unzipping potential from nonequilibrium data.} The data points in panels (a) show the average velocity and diffusion coefficient of the nanopore unzipping process at different forces, $f$. The continuous theoretical curves are based on Eqs.~\ref{eqn:GAD} and correspond to the least-square fit of the data via an optimal parametrization of the washboard potential, $U(x)$, of Eqs.~\ref{eqn:U}.
The potential is shown in panel (b) with and without various tilting forces.}
\label{fig3}
\end{figure}

The force dependence of the observed drift-diffusion regime is summarised in Fig.~\ref{fig3}, which presents the average unzipping velocity, $v$, and diffusion coefficient, $D$, computed respectively from the time-normalised mean and variance of the $\Delta n(t)$ traces (SM). Both observables vary considerably across the force range.

\textit{ One-dimensional model with tilted washboard potential.}
Restricting considerations to the drift-diffusion regime, it appears natural to postulate that the unzipping process
can be modeled as a stochastic process in a one-dimensional tilted washboard (periodic) potential. The assumption, which neglects factors such as the internal dynamics of the DNA chain, is suggested by the repetitive breaking of base pairs at the pore entrance. The latter, in fact, involves overcoming a free-energy barrier, after which the next upstream base pair is typically drawn at the pore and primed to undergo the same process.

For a stochastic process in a 1D washboard potential with periodicity $L$, $U(x)=U(x+L)$, and subject to an external force, $f_{\rm tilt}$, the expressions for the drift velocity and diffusion coefficient have been explicitly derived\cite{reimann2001giant,reimann2002} in connection with the giant acceleration of diffusion\cite{costantini1999threshold}:
\begin{equation}
\begin{cases}
\tilde{v}=\frac{1 - e^\frac{-L\cdot f_{\rm tilt}}{k_B\,T}}{\int_0^L dx\, I_{+}(x)/L}\ , \\
\\
\tilde{D}=D_0\frac{\int_0^L dx\, I^2_{+}(x) I_{-}(x)/L  }{[\int_0^L dx\, I_{+}(x)/L]^3}\ ,
\end{cases}
\label{eqn:GAD}
\end{equation}
\noindent where $I_{\pm}(x) = \int_0^L \frac{dy}{D_0} e^{[\pm U(x) \mp U (x \mp y) - y\, f_{\rm tilt}]/k_B T}$, $k_B T$ is the characteristic thermal energy, and $D_0$ is a nominal diffusion coefficient setting the overall scale of $\tilde{v}$ and $\tilde{D}$.

For our system, we identify the periodicity with the longitudinal spacing of the bases in the pore, $L=0.6$nm, and $x=L\, n$ with a cumulative longitudinal coordinate proportional to the number of translocated bases, $n$. The  previously introduced unzipping velocity, $v$, and diffusivity, $D$, can be related to those of Eq.~\ref{eqn:GAD} via $\tilde{v}=L\,v$ and $\tilde{D}=L^2\,D$. Finally, we put the driving and tilting forces in correspondence with $f_{\rm tilt} = f - f_0$.

With this assumed mapping, we analyzed the compliance of the unzipping data of Fig.~\ref{fig2} with the stochastic model of Eqs.~\ref{eqn:GAD}. The gist of the analysis was to derive the unknown effective periodic potential, $U(x)$, yielding the best match of the observed and theoretical data. To this end, we considered the following minimalistic parametrization for $U(x)$
\begin{equation}
  U(x) = \begin{cases}
 \frac{U_0}{2} \left [ \cos \left[ \frac{\pi (x_T -x)}{x_T} \right] +1 \right] \ & \text{for $ 0 \le x \le x_T$} \\
 \\
\frac{U_0}{2} \left [ \cos \left[ \frac{\pi (x -x_T)}{L-x_T} \right] +1 \right] \ & \text{for $ x_T \le x \le L$} .\\
\end{cases}
\label{eqn:U}
\end{equation}
\noindent The functional form provides a smooth (differentiable) periodic profile whose two essential features, namely the placement and height of the barrier, are respectively set by $x_T$ and $U_0$. These two quantities plus the $D_0$ coefficient, which sets the timescale, were treated as free parameters to be set by the least-square fit of the observed $v$ and $D$ data with those predicted theoretically from Eqs.~\ref{eqn:GAD}. Details of the fitting procedure and its robustness for alternative forms of the potential are provided in the SM.

The best-fitting $v$ and $D$ curves are shown as continuous lines in the plots of Fig.~\ref{fig3}a and provide a visibly good account of the observed data. We stress that the $v$ and $D$ datasets were fitted simultaneously, not separately. The visible agreement supports the assumption that the unzipping regime at intermediate forces can be modelled as a 1D normal diffusive process in a tilted periodic potential. This point is reinforced by the fact that the observed scaling of the translocation curves in the drift-diffusion regime is well accounted for by the force dependence of the velocities of Eq.~\ref{eqn:GAD}, see SM.

The baseline (untilted) periodic potential derived from the best-fit procedure is shown in Fig.~\ref{fig3}b. The barrier height is $U_0=18.0 \pm 1.5$pN\, nm, about 4.4$k_B\, T$ at 300K, while its location is  $x_T= 0.15 \pm 0.02$nm, yielding a noticeably asymmetric shape. The results are robust upon changing the functional form of $U(x)$ (SM). The asymmetry of the best-fit potential has direct bearings on the slope of the potential and thus on the tilting force required to suppress the barrier, see Fig.~\ref{fig3}b. Detailed calculations (SM) show that: (i) the critical tilting force yielding the maximum diffusion coefficient is about equal to 150pN, and (ii) that the Peclet number $Pe= \tilde{v}\, L/\tilde{D} = v/D$, remains of the order unity for $f_{\rm tilt} \lesssim 45$pN, equivalent to $f \lesssim 62$pN, and has a pronounced increase beyond it. The force above is comparable to $f_1\sim\textcolor{black}{60}$pN for which we observe the crossover from normal to anomalous drift-diffusion. The anomalous regime arguably originates from additional concurrent mechanisms caused by the growing relevance of advective transport over diffusion, such as the increasing strain of the DNA duplex approaching the pore entrance (SM).

\begin{figure}[b]
\centering 
\includegraphics[width=0.98\columnwidth]{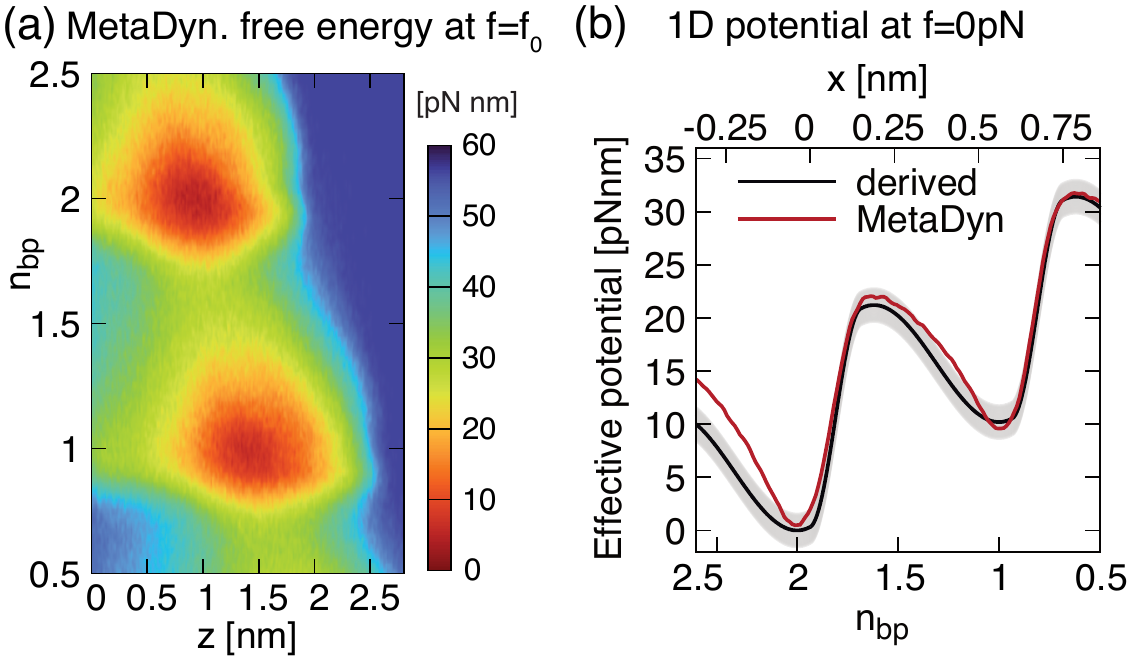}
\caption{{Comparison with unzipping free-energy from metadynamics calculations.} (a) Heatmap of the 2D free-energy landscape for unzipping of two consecutive base pairs at $f=f_0=17$pN. The landscape was computed with metadynamics simulations of oxDNA2, using two collective variables, $n_{bp}$ and $z$, see main text. (b) Thermodynamic reweighting and integration of the 2D landscape yield the potential in panel (b), which is the 1D free-energy profile at zero driving force, $f=0$. The profile (red) shows a good accord with the derived potential (black) tilted appropriately for zero driving force, $U(x)+f_0\, x$. The shaded band reflects the statistical uncertainty on $U_0$, 1.5pN nm. The linear correspondence of $x$ and $n_{bp}$ was set with the abscissa of the minima, which are displaced relative to the untilted case. The vertical offset of the potentials, which is immaterial for free-energy differences, was set by best matching the height of the two minima.}
\label{fig4}
\end{figure}

The key result is that the derived potential, $U(x)$, corresponds to the free-energy profile of the system at the stalling force, $f_0$. We established this by comparison with the thermodynamic potential computed with independent, conventional methods. Specifically, we used metadynamics \cite{laio2002escaping,micheletti2004reconstructing} simulations to obtain the translocation free-energy landscape for a short duplex segment, again modelled with oxDNA2. We reconstructed the landscape using two collective variables, the number of paired bases on the {\em cis} side, $n_{bp}$,  and the pore insertion depth, $z$, of the leading nucleotide of the segment, see SM. Fig.~\ref{fig4}a presents the landscape for two consecutive base pairs; the two minima associated with the paired states are readily identified. We then performed a thermodynamic reweighting and integration of the landscape (SM) to obtain the 1D unzipping free-energy profile at no applied force, $f=0$pN. The $n_{bp}$-dependent profile is shown in Fig.~\ref{fig4}b, where we overlaid the derived potential, again at no applied force ($U(x)+ f_0\, x$), with a linear correspondence of the reaction coordinates, $n_{bp}$ and $x$. The good agreement of the profiles is evident, the barrier heights difference being comparable to the estimated statistical error of $U_0$. The match establishes that $U(x)$ recovered from nonequilibrium trajectories is a \textit{bonafide} thermodynamic potential.

{\em Conclusions and outlook.} \textcolor{black}{We provided a systematic characterization and rationalization of DNA unzipping by nanopore translocation, a fundamental process from \textit{ in vivo} biological systems to DNA sequencing.} Further, we recovered the detailed thermodynamics of single base pair breaking using only nonequilibrium trajectories. This is made possible by the stochastic repetition of the elementary unzipping steps in the translocation of a long duplex, whose thermodynamics can then be unravelled with the theory of stochastic processes in tilted periodic potentials. In this regard, our strategy complements other schemes that provided breakthrough demonstrations that extracting single-molecule thermodynamics from driven processes is feasible\cite{jarzynski1997nonequilibrium,crooks1999entropy,collin2005verification,dudko2006,dudko2007,Dudko2008,dudko2013,rissone2022stem}. \textcolor{black}{These include the theory of driven diffusive barrier crossings for the cooperative unfolding of DNA and RNA motifs\cite{dudko2007,dudko2008theory,gupta2011experimental}, as well as the recovery of sequence-dependent interactions by mechanically unzipping nucleic acids of hundreds of base pairs\cite{huguet2010single,huguet2017derivation}.}

\textcolor{black}{Our approach is expectedly applicable in broader single-molecule contexts involving tilted periodic potentials. These include molecular motors under external torque\cite{hayashi2015giant} and the transport of filamentous molecules across regular arrays of obstacles\cite{kim2017giant}, whose giant acceleration of diffusion was shown to comply with eqs.~\ref{eqn:GAD} with known or postulated potentials.} The possibility to derive the potential or free-energy profile from the data itself, rather than prior knowledge, would substantially broaden the range of addressable systems and would allow extracting properties of the barriers or transition states not otherwise obtainable from nonequilibrium trajectories.

\bibliography{bibliography.bib}

\end{document}